\documentclass[12pt]{article}
\usepackage[margin=1in]{geometry}
\usepackage{microtype}
\usepackage{amsmath}
\usepackage{amssymb}
\usepackage{amsfonts}
\usepackage{bm}
\usepackage{color,soul,url}
\usepackage{subfig}
\usepackage{natbib}
\newtheorem{theorem}{Theorem}[section]

\newtheorem{example}{Example}
\newtheorem{remark}{Remark}[theorem]
\newtheorem{assumption}{Assumption}

\usepackage{enumerate}

\usepackage{enumitem} 
\newlist{condenum}{enumerate}{1} 
\setlist[condenum]{label=(C\arabic*), 
	ref=(C\arabic*), wide}

\usepackage{mathtools}
\DeclarePairedDelimiter\abs{\lvert}{\rvert}%
\DeclarePairedDelimiter\norm{\lVert}{\rVert}%
\makeatletter
\let\oldabs\abs
\def\abs{\@ifstar{\oldabs}{\oldabs*}}
\let\oldnorm\norm
\def\norm{\@ifstar{\oldnorm}{\oldnorm*}}
\makeatother

\newcommand{\mD}{\mathcal{D}}

\def\transpose{^{\sf \scriptscriptstyle{T}}}
\newcommand{\ind}{\perp\!\!\!\perp}

\newcommand{\inv}{^{-1} }
\def\dd{\mathcal{D}}
\def\ddd{{O}}

\def\mystar{*}%\star
\def\ppsi{\psi}
\def\ppsi{U}

\title{Online Causal Inference with Application to Near Real-Time Post-Market Vaccine Safety Surveillance}

\author{Xu Shi\thanks{Department of Biostatistics, University of Michigan} ~and~Lan Luo\thanks{Department of Statistics and Actuarial Science, University of Iowa}}
\date{}

\begin{document}
	
	\maketitle
	
	\begin{abstract}
	Streaming data routinely generated by mobile phones, social networks, e-commerce, and electronic health records present new opportunities for near real-time surveillance of the impact of an intervention on an outcome of interest via causal inference methods. However, as data grow rapidly in volume and velocity, storing and combing data become increasingly challenging.  The amount of time and effort spent to update analyses can grow exponentially,  which defeats the purpose of instantaneous surveillance.   Data sharing barriers in multi-center studies bring additional challenges to rapid signal detection and update.  It is thus time to turn static causal inference to online causal learning that can incorporate new information as it becomes available without revisiting prior observations.  In this paper, we present a framework for online estimation and inference of treatment effects leveraging a series of datasets that arrive sequentially without storing or re-accessing individual-level raw data.   We establish estimation consistency and asymptotic normality of the proposed framework for online causal inference.  In particular, our framework is robust to biased data batches in the sense that the proposed online estimator is asymptotically unbiased as long as the pooled data is a random sample of the target population regardless of whether each data batch is.  We also provide an R package for analyzing streaming observational data that enjoys great computation efficiency compared to existing software packages for offline analyses. Our proposed methods are illustrated with extensive simulations and an application to sequential monitoring of adverse events post COVID-19 vaccine.
	\end{abstract}
	
	\section{Introduction}
	The growing availability of massive streaming data assembled through, for example, mobile or web applications%~\cite[]{Mobile_users_2018}
	, e-commerce purchases%~\cite[]{ecommerce_2016}
	, infectious disease surveillance programs %~\cite[]{disease2016,disease2020}, 
	and administrative healthcare databases, is opening up new research opportunities but also comes with challenges. First, it is time consuming and resource intensive to store and maintain massive amount of data with possibly unbounded size. Second, it is a large and seemingly unnecessary workload to sequentially combine existing data with a newly arrived batch of data and re-analyze an updated dataset. 
	Third, in multi-center studies, it is challenging and potentially infeasible to continuously pool data across different locations and time due to privacy concerns.

	An important unmet need is to process such rapidly growing streams of data as they become available sequentially and perpetually over time without storing or revisiting prior observations. 
	Ongoing efforts have been made to develop modern distributed computing frameworks such as Hadoop, Spark and Storm~\cite[]{Hadoop2003,MapReduce2004,Spark2015,Storm_Spark2016}. Aligned with these architectures, various statistical methods and algorithms for online estimation and inference have been proposed, including aggregated estimating equation~\cite[]{lin2011aggregated}, stochastic gradient descent~\cite[]{Robbins1951,Sakrison1965,Toulis2015MLE}, cumulative estimating equation~\cite[]{Schifano2016CUEE}, as well as renewable estimator~\cite[]{Luo2020}.

	However, the majority of research efforts are centered around associative and predictive studies rather than providing insights into the impact of an intervention on an outcome of interest via causal inference. 
	In this paper, 
	we propose a novel framework for online causal learning in the point exposure setting that can incorporate new information as it becomes available without storing or revisiting prior observations. We address the following questions: (i) how to utilize observational streaming data to make causal inference? (ii) what types of summary statistics to be stored and updated without revisiting prior observations; and (iii) whether one could achieve similar statistical properties to the offline causal inference methods which require combining raw data streams. 
	We first introduce a motivating example and then discuss the statistical contributions of this paper.
	
	\subsection{Safety monitoring of medical products}
	Turning static causal inference to real-time causal learning is particularly important for post-market safety monitoring of medical products including drug, vaccine, and medical devices using routinely collected administrative data such as claims and electronic health records data \citep{nelson2015methods}. 
	Pre-approval clinical trials are often too small to systematically detect rare adverse events and may not be generalizable to the population who will ultimately receive the medical products post-approval.
	Therefore, the Food and Drug Administration’s (FDA) created the Sentinel Initiative using administrative databases from numerous large health plans across the US with approximately 193 million patients to monitor the safety of all regulated medical products after they have been posted to the market. 
	Similarly, the Centers for Disease Control and Prevention (CDC) created the Vaccine Safety Datalink to evaluate the safety of vaccines when there is a new vaccine recommended for use or a change in how a vaccine is recommended. Active surveillance systems using routinely collected data streams from large, population-based cohorts is powerful as it represents a broader population than typically enrolled in clinical trials, and it enables early detection of safety signals for less common adverse outcomes.
	
	Existing statistical methods for active monitoring includes disproportionality analyses \citep{moore1997reports,bate1998bayesian,evans2001use}, continuous sequential testing \citep{wald1945sequential,page1954continuous,grigg2003use,brown2007early,brown2009early}, and group sequential testing \citep{whitehead1997design,jennison1999group,li2011propensity,cook2012statistical}. Specifically, the disproportionality measure is a ratio of the observed and expected numbers of adverse event among a specific drug user group of interest, where the expected number is calculated based on the null hypothesis of no association between the drug and the adverse event in view. The continuous and group sequential testing methods, originated from randomized clinical trials and applied in observational study settings, evaluate safety signals as data are accumulated with a pre-defined stopping rule, and stops further sampling as soon as significant results are reached at a potentially much earlier time.
	However, as new data become available, the above methods require one to combine new data with all prior observations to update the analysis. In addition,  confounding adjustment is often limited to stratification due to patient privacy concerns.  
	
	\subsection{Statistical contributions}
	In this paper, we propose a general framework for online estimation and inference of the causal effect of an intervention on an outcome using streams of observational data that arrive over a series of time points. Our proposal is motivated by the observation that for most commonly used causal inference methods, one can jointly estimate both nuisance parameters and the treatment effect of interest by stacking estimating equations for individual parameters and solving a vector estimating equation, often referred to as M-estimation \citep{stefanski2002calculus}. With streaming data, instead of solving an \textit{offline estimating equation} evaluated on the aggregated data from all batches, one can solve an \textit{online estimating equation} which approximates its offline counterpart and allows one to incrementally update the estimator and the variance-covariance matrix, when a new batch of data arrives without revisiting previous batches. Moreover, although each data batch may be a biased sample irrepresentative of the target population, we show that our online estimator is asymptotically unbiased as long as the pooled data is a random sample of the target population.  
	
	 Our statistical contributions include: (i) under mild regularity conditions, the proposed online estimator is asymptotically equivalent to the offline oracle estimator obtained by pooling all raw data, (ii) it is robust to biased data batches, as long as the pooled data is representative of the target population (iii) it allows synchronization of evidence regardless of the sample size of newly arrived dataset; and (iv) it only requires storing and integration of summary statistics, and thus enjoys great computational efficiency compared to the offline method and can be applied to overcome data sharing barrier. It is also worth noting that this framework encompasses a wide range of causal inference settings including mediation analysis \citep{robins1992identifiability,pearl2001proceedings,tchetgen2012semiparametric}, instrumental variable methods \citep{angrist1996identification,hernan2006instruments}, and proximal causal inference \citep{tchetgen2020introduction,cui2020semiparametric}. To facilitate real-world application of sequential monitoring of drug safety using observational streaming data, we further propose a sequential testing framework. 
	
	 The paper is organized as follows. In Section~\ref{sec:methods} we first review several classic offline causal inference methods, and then present our online framework. We establish its large sample properties and propose a sequential testing method. In Section~\ref{sec:sim} we conduct simulation experiments  to evaluate the performance of our proposed method in comparison to offline benchmarks. In Section~\ref{sec:application}, we apply  online causal inference methods to monitor adverse events associated with COVID-19 vaccine using data from the Vaccine Adverse Event Reporting System (VAERS). Finally, we make concluding remarks in Section~\ref{sec:discuss}. Extensions of our strategy to a wide class of causal inference methods as well as a detailed proof of the large sample properties are presented in the appendix.

	\section{Methods\label{sec:methods}}
	\subsection{Review of commonly-used offline causal inference methods\label{subsec:offline}}
	Let $A\in\{0,1\}$ denote the binary treatment, $Y\in\mathcal{R}$ denote the outcome of interest, and $X\in\mathcal{R}^p$ denote the pre-treatment covariates with the first component being one. 
	We briefly review a classical causal inference methods, which are applicable to a single, static dataset, referred to as offline data: 
	\begin{assumption}[Offline data]\label{assump:obs}
		We observe a simple random sample from the target population with i.i.d.
		observations ${\ddd}_i =\{Y_i, A_i, X_i\transpose\}\transpose,
		i=1,\dots,n$.
	\end{assumption}
	We adopt the potential outcomes framework \citep{neyman1923applications,rubin1974estimating} to define causal effects. Let $Y(a)$, $a=0$ or $1$, denote the potential outcome had the subject, possibly contrary to the fact, been given treatment $a$. 
	Without loss of generality, we consider the average treatment effect (ATE) defined as $\Delta=\mathbb{E}[Y(1)-Y(0)]$,
	and discuss extensions to other causal estimands in Appendix~A.1. %\ref{sec:extend}.
	A fundamental problem in identifying the ATE is that for each subject, only one potential outcome is ever observed.  
	One seeks minimal assumptions to identify the ATE. Following \cite{rosenbaum1983central}, we make the following assumptions to identify the ATE. 
	\begin{assumption}[Consistency and positivity]\label{assump:cons_pos}
		Consistency: $Y(a) =Y$ almost surely when $A=a$; Positivity: $0< \mathbb{P}(A=a|X)<1$ for all $a$ almost surely.
	\end{assumption}
	The consistency assumption ensures that the exposure is defined with enough specificity such that among people with $A=a$, the observed outcome $Y$ is a realization of the potential outcome value $Y(a)$. The positivity assumption states that in all observed covariate strata there are always some individuals with treatment value $A=a$ for all $a$.
	\begin{assumption}[No unmeasured confounding]
		$(Y(0),Y(1))\ind A\mid X$.\label{assump:ignorability}
	\end{assumption}
	Assumption~\ref{assump:ignorability} states that the treatment assignment is ignorable given $X$, which rules out the existence of unmeasured confounding. 
	There are a range of methods to deal with unmeasured confounding such as the instrumental variable method \citep{angrist1991does,angrist1996identification,hernan2006instruments} and the proximal causal inference method \citep{miao2018identifying,shi2020multiply,tchetgen2020introduction,cui2020semiparametric}. Extension of  adjustment for unmeasured confounding to online setting will be discussed in Appendix~A.1. %\ref{sec:extend}. 
	Under Assumptions~\ref{assump:obs}-\ref{assump:ignorability}, the ATE is identified by 
	\begin{equation}
	    \Delta=
	\int_{\mathcal{X}}
	\{\mathbb{E}[Y\mid A=1,X=x]-\mathbb{E}[Y\mid A=0,X=x]\}
	f(x)dx.\label{eq:ate}
	\end{equation}
	The ATE is a functional of the distribution of ${\ddd}$, which is often unknown and need to be modeled and estimated. To this end, we introduce the following working models for propensity score (PS) and outcome regression (OR):
	\begin{assumption}[Propensity score model]\label{assump:PS}
		The conditional treatment probability is correctly specified as $\mathbb{P}(A=1\mid X)=e(X; \alpha)$, indexed by finite-dimensional parameter $\alpha$.
	\end{assumption}
	\begin{assumption}[Outcome regression model]\label{assump:OR}
		The conditional outcome mean model is correctly specified as $\mathbb{E}[Y\mid A,X]=m(A,X; \beta)$, indexed by finite-dimensional parameter $\beta$.
	\end{assumption}
	
	Both PS and OR models are nuisance parameters that are not of scientific interest but are fitted in order to estimate the ATE.
	Common practice is to first fit the PS and/or OR models, then plug in the estimated model parameters  to a formula that produces an estimator for the ATE. For example, the inverse probability of treatment weighting method is a two-stage approach: one first estimates $\alpha$ in a pre-specified propensity score model by solving $\sum_{i=1}^n {g}(X_i)[A_i - e(X_i;{\alpha})]=0$, 
	then plugs in $\widehat{\alpha}$ to $\sum_{i=1}^n \frac{A_iY_i}{e(X_i;{{\alpha}})} -
	\frac{(1-A_i)Y_i}{1-e(X_i;{{\alpha}})} 	-\Delta=0$ 
	to obtain an estimator of the ATE $\widehat{\Delta}$.

	An important observation is that, instead of such a multi-stage estimation procedure, one can stack estimating functions for estimation of all necessary parameters and estimate them jointly as the solution to a vector estimating equation
	\begin{equation}
		\sum_{i=1}^{n} {\ppsi}({\ddd}_i;{\theta})=0,\label{eq:mest}
	\end{equation}
	where $\theta\in\mathcal{R}^p$ denotes all necessary parameters, and ${\ppsi}({\ddd}_i;{\theta})$ is a vector of estimating functions. 
	Let $\theta_0$ denotes the true value,  we present the following examples.
	\begin{example}[Inverse probability of treatment weighting] The inverse probability of treatment weighting (IPTW) estimator can be obtained as a solution to Eq.~\eqref{eq:mest} with
		\[
		{\ppsi}({\ddd};{\theta})= 
		\begin{pmatrix}
			{g}(X)[A - e(X;{\alpha})] \\
			\frac{AY}{e(X;{\alpha})} -
			\frac{(1-A)Y}{1-e(X;{\alpha})} 	-\Delta
		\end{pmatrix},
		\]
		where ${\theta}=({\alpha}\transpose,\Delta)\transpose$, ${g}(X)$ is a user specified function of the same dimension as $\alpha$. 
		Under Assumptions~\ref{assump:cons_pos}-\ref{assump:PS},  we have  $\mathbb{E}_{{\theta}}\left[{\ppsi}(O_i,{\theta}) \right] = 0$ if and only if ${\theta} ={\theta}_0$.
	\end{example}
	\begin{example}[G-computation] The G-computation (G-comp) estimator can be obtained as a solution to Eq.~\eqref{eq:mest} with
		\[
		{\ppsi}({\ddd};{\theta})= 
		\begin{pmatrix}
			{h}(A,X)[Y - m(A,X;{\beta})] \\
			m(1,X;{\beta}) - m(0,X;{\beta})
			-\Delta
		\end{pmatrix},
		\]
		where ${\theta}=(\beta\transpose,\Delta)\transpose$,  ${h}(A,X)$ is a user specified function of the same dimension as $\beta$. 
		Under Assumptions~\ref{assump:cons_pos}-\ref{assump:ignorability} and \ref{assump:OR}, we have  $\mathbb{E}_{{\theta}}\left[{\ppsi}(O_i,{\theta}) \right] = 0$ if and only if ${\theta} ={\theta}_0$.
	\end{example}
	\begin{example}[Augmented inverse probability of treatment weighting] The augmented inverse probability of treatment weighting (AIPTW) estimator can be obtained as a solution to Eq.~\eqref{eq:mest} with
		\[
		{\ppsi}({\ddd};{\theta})= 
		\begin{pmatrix}
			{g}(X)[A - e(X;{\alpha})] \\
			{h}(A,X)[Y - m(A,X;{\beta})] \\
			\left[ m(1,X;{\beta})+\frac{A[Y-m(1,X;{\beta})]}{e(X;{\alpha})}  \right] -
			\left[
			m(0,X;{\beta})+\frac{(1-A)[Y-m(0,X;{\beta})]}{1-e(X;{\alpha})} 
			\right]	-\Delta
		\end{pmatrix},
		\]
		where ${\theta}=({\alpha}\transpose,{\beta}\transpose,\Delta)\transpose$. If Assumptions~\ref{assump:cons_pos}-\ref{assump:ignorability} hold, and either Assumption~\ref{assump:PS} or Assumption~\ref{assump:OR} holds, we have $\mathbb{E}_{{\theta}}\left[{\ppsi}(O_i,{\theta}) \right] = 0$ if and only if ${\theta} ={\theta}_0$.
	\end{example}
	In fact, as we further illustrate in Section~A.1 %\ref{sec:extend}
	of the Appendix, a large number of causal inference methods can be written in the form of an M-estimator $\widehat{\theta}$ which solves  $\sum_{i=1}^nU(O_i;\widehat{\theta})=0$   \citep{stefanski2002calculus}.
	The M-estimation is first introduced in  \cite{godambe1960optimum}, \cite{huber1964robust} and \cite{huber1967under}, and has been extended to longitudinal setting as generalized estimating equations (GEE) by \cite{liang1986longitudinal}. A key advantage of formulating causal inference methods as M-estimators is to allow for convenient computation and inference  via a unified large sample approximation. In particular, under suitable regularity conditions, the M-estimator $\widehat{\theta}$ is consistent and asymptotically normal \citep{godambe1960optimum,huber1964robust,huber1967under}
	\[
	\sqrt{n}(\widehat{\theta}-\theta_0) \overset{d}{\to} \mathcal{N}(0, \mathbb{V}(\theta_0)), \ \text{as}\ n\to\infty
	\]
	where the asymptotic variance of all parameters is given by
	$\mathbb{V}(\theta_0)=\mathbb{S}(\theta_0)\inv \mathbb{M}(\theta_0)\left\{\mathbb{S}(\theta_0)\right\}\transpose $, with $\mathbb{S}(\theta_0)=\mathbb{E}[-\partial U(O_1;\theta)/\partial\theta\transpose \mid _{\theta=\theta_0} ]$, referred to as the sensitivity matrix, and $\mathbb{M}(\theta_0)=\mathbb{E}[U(O_1;\theta_0)U(O_1;\theta_0)\transpose ]$, referred to as the variability matrix.

	\subsection{A general strategy for online causal inference}\label{ssec:online_estimator}
	We present a general strategy for online causal inference by extending M-estimation from offline to online setting to allow for sequential update of causal effect as new data become available without revisiting prior observations. Our strategy is straightforward: we summarize information from previous batches, and when a new batch of data arrives, we combine the new data with the summary statistics from old data to construct an {\em online estimating equation}, from which we solve for an updated estimator and incrementally update the variance-covariance matrix.
	The rational is that the summary statistics from previous batches form an empirical ``prior" of $\theta$, which is then combined with the current batch of data to form a ``posterior"  of $\theta$ \citep{efron1993bayes}.
	
	Suppose data arrive in a series of data batches, 
	denoted by 
	${\dd}_1=\{\ddd_1,\dots,\ddd_{n_1}\}\transpose$, ${\dd}_2=\{\ddd_{n_1+1},\dots,\ddd_{n_1+n_2}\}\transpose$, $\dots$, ${\dd}_b=\{\ddd_{n_{b-1}+1},\dots,\ddd_{n_{b-1}+n_b}\}\transpose$, $\dots$, 
	where $n_j$ is the sample size of the $j$-th data batch.
	Traditionally, when a new batch ${\dd}_b$ arrives, one would pull data from all $b$ batches to re-estimate $\theta$ by solving $\sum_{i=1}^{N_b}\ppsi(\theta;\ddd_{i})=0$, 
	where $N_b=\sum_{j=1}^bn_j$ is aggregated sample size up to the $b$-th batch.
	In the era of big data, $N_b$ can be extremely large, making it computationally intractable to re-estimate based on the aggregated data. Our goal is to update estimates without using any subject-level data but only summary statistics from previous batches of data. 
	
	To illustrate our method, we begin with two data batches. Let \[{U}_j(\dd_j;\theta)=\sum_{i\in\dd_j}\ppsi(\ddd_{i};\theta)\] denote the sample sum of the estimating function evaluated at the $j$-th batch of data $\dd_j$. Note that with slight abuse of notation we also use $\dd_j$ to denote the set of indices for observations in this batch.
	Let ${\widehat{\theta}}_1^\mystar$ be the initial estimate obtained from the first batch ${\dd}_1$, which solves
	\begin{equation}
		{\ppsi}_1({\dd}_1;{\widehat{\theta}}_1^\mystar)={0}, \label{eq:firstbatch}
	\end{equation}
	When the second data batch ${\dd}_2$ arrives, traditional offline estimator solves the following {\it aggregated estimating equation} 
	\begin{equation}\label{eq:2batch}
		{\ppsi}_1({\dd}_1;{\widehat{\theta}}_2^\mystar) + {\ppsi}_2({\dd}_2;{\widehat{\theta}}_2^\mystar) = {0}.
	\end{equation}
	However, the first term ${\ppsi}_1({\dd}_1;{\widehat{\theta}}_2^\mystar)$ requires revisiting the first data batch ${\dd}_1$ and evaluating ${U}_1(\dd_1;\theta)$ at a range of $\theta$ values to solve Eq.~\eqref{eq:2batch} for ${\widehat{\theta}}_2^\mystar$. 
	To derive a renewable estimation that does not revisit ${\dd}_1$, we propose an online estimator ${\widetilde{\theta}}_2$ that solves the following {\it online estimating equation}
	\begin{equation}\label{eq:2_batch_ee}
		{S}_1({\dd}_1;{\widehat{\theta}}_1^\mystar) ({\widehat{\theta}}_1^\mystar - {\widetilde{\theta}}_2) + {\ppsi}_2({\dd}_2;{\widetilde{\theta}}_2) = {0},
	\end{equation}
	where ${S}_1({\dd}_1;{\widehat{\theta}}_1^\mystar) =-{\partial {\ppsi}_1({\dd}_1;\theta)}/{\partial\theta\transpose} \mid _{\theta ={\widehat{\theta}}_1^\mystar}$. 
	Through~\eqref{eq:2_batch_ee}, the initial estimate ${\widehat{\theta}}_1^\mystar$ is updated to ${\widetilde{\theta}}_2$. 
	Importantly, the updating only requires summary statistics, ${S}_1({\dd}_1;{\widehat{\theta}}_1^\mystar)$, and the old estimate, $\widehat{\theta}_1$, from the first data batch $\mathcal{D}_1$, without requiring any subject-level data. 
	
	Our proposal is motivated by the following observation. 
	Taking the first-order Taylor expansion of ${\ppsi}_1({\dd}_1;{\widehat{\theta}}_2^\mystar)$ around ${\widehat{\theta}}_1^\mystar$ we have
	\begin{equation}\label{eq:2batch_taylor}
		\begin{split}
			{\ppsi}_1({\dd}_1;{\widehat{\theta}}_2^\mystar)=&~
			{\ppsi}_1({\dd}_1;{\widehat{\theta}}_1^\mystar) + {S}_1({\dd}_1;{\widehat{\theta}}_1^\mystar)({\widehat{\theta}}_1^\mystar-{\widehat{\theta}}_2^\mystar) + 
			O_p\left(\frac{n_1}{N_2}\| {\widehat{\theta}}_2^\mystar - {\widehat{\theta}}_1^\mystar\|^2\right)\\
			\approx&~{S}_1({\dd}_1;{\widehat{\theta}}_1^\mystar)({\widehat{\theta}}_1^\mystar-{\widehat{\theta}}_2^\mystar),
		\end{split}
	\end{equation}
	where the approximation holds because ${\ppsi}_1({\dd}_1;{\widehat{\theta}}_1^\mystar) ={0}$ by \eqref{eq:firstbatch} and the higher order error term $O_p\left({n_1}{N_2}\inv \|{\widehat{\theta}}_2^\mystar-{\widehat{\theta}}_1^\mystar \|^2\right)$ may be asymptotically ignored when $N_2\to\infty$. 
	Therefore, ${\ppsi}_1({\dd}_1;{\widehat{\theta}}_2^\mystar)$ can be approximated using ${S}_1({\dd}_1;{\widehat{\theta}}_1^\mystar)$ and ${\widehat{\theta}}_1^\mystar$, which are summary statistics of $\dd_1$ that do not involve pooling of individual-level data.

	We now formally introduce our proposed online causal inference method for sequentially updated estimation and inference. 
	Generalizing Eq.~\eqref{eq:2_batch_ee} to streaming data that arrive in a series of batches, we have the  following online estimating equation  upon the arrival of data batch $\dd_b$:
	\begin{equation}\label{eq:b_batch_ee}
		{\widetilde{\ppsi}}_b ({\widetilde{\theta}}_b)= \sum_{j=1}^{b-1}{S}_j({\dd}_j;{\widetilde{\theta}}_j)({\widetilde{\theta}}_{b-1}-{\widetilde{\theta}}_b) + {\ppsi}_b({\dd}_b;{\widetilde{\theta}}_b) = {0},
	\end{equation}
	where ${\widetilde{\theta}}_1={\widehat{\theta}}_1^\mystar$ at the initial data batch ${\dd}_1$. 
	Eq.~\eqref{eq:b_batch_ee} combines the current data batch ${\dd}_b$ and summary statistics $\left\{{\widetilde{\theta}}_{b-1},{\widetilde{S}}_{b-1} \right\}$ from historical data. 
	Solving Eq.~\eqref{eq:b_batch_ee} may be implemented via an iterative algorithm such as the Newton-Raphson algorithm:
	\begin{equation}\label{eq:algorithm}
		{\widetilde{\theta}}_b^{(r+1)} = {\widetilde{\theta}}_b^{(r)} + 
		\left\{{\widetilde{S}}_{b-1} + {S}_b({\dd}_b;{\widetilde{\theta}}_{b}^{(r)}) \right\}\inv  {\widetilde{\ppsi}}_b^{(r)},
	\end{equation}
	where ${\widetilde{\ppsi}}_b^{(r)} = {\widetilde{S}}_{b-1}\left({\widetilde{\theta}}_{b-1} - {\widetilde{\theta}}_b^{(r)}\right) + {\ppsi}_b\left({\dd}_b;{\widetilde{\theta}}_b^{(r)}\right)$ denotes the adjusted estimating function and ${\widetilde{S}}_{b-1} = \sum_{j=1}^{b-1}{S}_j({\dd}_j;{\widetilde{\theta}}_j)$. 
	For inference, we use an incrementally updated variance-covariance matrix given by
	\begin{equation}\label{eq:var_theta}
		\widetilde{\text{var}}(\widetilde{{\theta}}_b) = \widetilde{{S}}_b\inv \widetilde{{M}}_b \left(\widetilde{{S}}_b\transpose\right)\inv ,
	\end{equation}
	where $\widetilde{M}_b=\sum_{j=1}^b \sum_{i\in\mathcal{D}_j}{U}(O_{i};\widetilde{\theta}_j){U}(O_{i};\widetilde{\theta}_j)\transpose $. Both $\widetilde{S}_b$ and $\widetilde{M}_b$ can be updated upon the arrival of a new data batch and stored for future aggregation. Suppose the length of $\widetilde{{\theta}}_b$ is $d$, then by definition of $U(O,\theta)$, the online estimator $\widetilde{\Delta}_b$ of the ATE is the last entry of $\widetilde{{\theta}}_b$ and its variance is given by the $(d,d)$-th entry of $\widetilde{\text{var}}(\widetilde{{\theta}}_b)$, i.e. $\widetilde{\text{var}}(\widetilde{\Delta}_b)=[\widetilde{\text{var}}(\widetilde{{\theta}}_b)]_{d,d}$.

	\subsection{Large sample properties of online estimator}\label{ssec:thm}
	We establish large sample properties for our proposed online estimator $\widetilde{\theta}_b$ in Eq.~\eqref{eq:b_batch_ee} under the regime where  $n_j$ is finite for every $j=1,\dots,b$, while the number of data batches arriving over time $b\to\infty$ such that $N_b=\sum_{j=1}^{b}n_j\to\infty$. The technical difficulty arises from the fact that $n_j$ is finite and the convergence of online estimator is driven by the number of iterative steps indexed by $b$. This theoretical framework is well aligned with streaming data collection scheme where data batches with finite sample sizes arrive perpetually over time. We have the following theorems which are proved in Sections~A.3-A.5 %\ref{proof:consistency}-\ref{proof:diff_U} 
	of the Appendix.
	
	\begin{theorem}\label{thm:consist}
		Under  Assumptions~\ref{assump:obs}-\ref{assump:OR} and regularity conditions stated in Section~A.2 %\ref{subsec:regular}
		of the Appendix, the proposed online estimator $\widetilde{\theta}_b$ in Eq.~\eqref{eq:b_batch_ee} is consistent
		\[
		\widetilde{\theta}_b \overset{p}{\to} \theta_0, \ \text{as} \ N_b=\sum_{j=1}^{b}n_j\to\infty.
		\] 
	\end{theorem}
	\begin{remark}
		It follows directly from Theorem~\ref{thm:consist} that our proposed online estimator for ATE is consistent, i.e. $\widetilde{\Delta}_b \overset{d}{\to}\Delta_0$, as $N_b\to\infty$.
	\end{remark}
	
	\begin{theorem}\label{thm:normal}
		Under  Assumptions~\ref{assump:obs}-\ref{assump:OR} and regularity conditions stated in Section~A.2 %\ref{subsec:regular}
		of the Appendix, the online estimator $\widetilde{\theta}_b$ given in Eq.~\eqref{eq:b_batch_ee} is asymptotically normally distributed, that is,
		\[
		\sqrt{N_b}(\widetilde{\theta}_b - \theta_0) \overset{d}{\to} \mathcal{N}\left\{{0},\mathbb{V}(\theta_0)\right\}, \ \text{as} \ N_b=\sum_{j=1}^{b} n_j\to\infty,
		\]
		where the asymptotic covariance matrix $\mathbb{V}(\theta_0) = \mathbb{S}(\theta_0)\inv \mathbb{M}(\theta_0)\left\{\mathbb{S}(\theta_0)\inv \right\}\transpose$. 
	\end{theorem}
	\begin{remark}
		By Theorem~\ref{thm:normal}, we have $\sqrt{N_b}(\widetilde{\Delta}_b - \Delta_0)\overset{d}{\to}\mathcal{N}(0, [\mathbb{V}(\theta_0)]_{d,d})$, as $N_b\to\infty$, where $[\mathbb{V}(\theta_0)]_{d,d}$ is the $(d,d)$-th entry of the asymptotic variance in Theorem~\ref{thm:normal}.
	\end{remark}
	
	We highlight that when analyzing the very first batch of data, we do not require its sample size $n_1$ to be large (which is unrealistic in online settings), and we only require the sample sensitivity matrix to be non-singular, which implicitly requires that $n_1$ is greater than the number of parameters $p$.

	\begin{theorem}\label{thm:diff_EE}
	Let $U_b^\star({\theta})=\sum_{j=1}^bU_j(\mD_j;\theta)$ denote the aggregated offline estimating equation, and $\widehat{\theta}_b^\star$ is the offline estimator satisfying $U_b^\star(\widehat{\theta}_b^\star)=0$. Under Assumptions~\ref{assump:obs}-\ref{assump:OR} and regularity conditions stated in Section~A.2 %\ref{subsec:regular} 
	of the Appendix, our proposed online estimator $\widetilde{\theta}_b$ given in Eq.~\eqref{eq:b_batch_ee} satisfies $\frac{1}{N_b}U_b^\star(\widetilde{\theta}_b)\overset{p}{\to}0$ as $N_b\to\infty$.
	\end{theorem}
	
	This theorem shows the proposed online estimator $\widetilde{\theta}_b$ is asymptotically unbiased. In practice, since each batch comes with a finite sample size, it may constitute a subpopulation or strata of the entire population, which is a biased sample in the sense that it is not representative of the target population. For example, in post-marketing surveillance of vaccine and drug safety, earlier data batches may be collected from individuals who need the medication the most, such as seniors, while later data batches may consist of more younger people. Although as $N_b\to\infty$, the pooled sample is representative of the target population, each data batch is a biased sample. 
	In particular, we consider a case where the distribution of the pre-treatment covariates $X$ in batch $b$ is not representative of the target population. That is, the selection mechanism from the aggregated data into batch $b$ to depend on $X$.  Consequently, even though the unbiasedness of the estimating equation holds over the entire population, it may be violated for a certain batch, i.e. $\mathbb{E}[U_b(\mD_b;\theta_0)]\neq 0$ where the expectation is taken over the distribution of this subpopulation. This theoretical results implies that our proposed online updating procedure is robust to such sampling variability and the resulting estimator $\widetilde{\theta}_b$ is asymptotically the root to the oracle offline estimating equation which is unbiased.
	
	\subsection{Sequential testing}\label{ssec:seq_test}
	Given that the sequentially updated estimator of ATE and its standard error are readily available as derived in Section~\ref{ssec:thm}, we can construct a sequential testing framework based on the Wald test as a way to sequentially monitor drug safety using observational streaming data. First, we define the following hypothesis of interest
	\[
	H_0:\Delta = \Delta_0 \ \text{v.s.} \ H_1: \Delta\neq \Delta_0,
	\]
	where $\Delta_0$ is a postulated true ATE. A Wald test statistic based on $\widetilde{\Delta}_b$ and $\widetilde{\text{var}}(\widetilde{\Delta}_b)$ is given by
	\begin{equation}\label{eq:Wald}
		{Z}_b  = \frac{\widetilde{\Delta}_b - \Delta_0}{\sqrt{\widetilde{\text{var}}(\widetilde{\Delta}_b)}}.
	\end{equation}
	
	To control the overall type I error, we use the alpha-spending function approach~\citep{Lan1983}. With a pre-fixed value $T$ of the total number of analyses, let $\alpha(t)$ denote the cumulative amount of type I error up to the $t$-th interim analysis, we assume that $0 < \alpha(1)\leq\dots\leq\alpha(T)=\alpha$, where $\alpha$ is the desired overall type I error to spend across all $T$ analyses. Boundary values $z(t)$ can be determined successively so that
	\[
	P_{\Delta_0}\left\{ \abs{{Z}_1}\geq z(1),\text{or} \ \abs{{Z}_2}\geq z(2), \text{or}, \dots, \text{or}\ \abs{{Z}_t}\geq z(t) \right\} = \alpha(t).
	\]
	The most commonly used boundary function for safety monitoring has been the Pocock boundary function $\alpha(t)=\alpha\log(1 +(e-1)t/T )$~\citep{Pocock1982}. It is almost flat and spends $\alpha$ approximately evenly across analyses. Another widely used function in efficacy studies is the O'Brien Fleming boundary $\alpha(t) = 2\left(1 - \Phi(z_{1- \alpha/2}/\sqrt{t/T}) \right)$ where $\Phi(\cdot)$ is the cumulative distribution function of the standard normal distribution~\citep{OBrien1979}.

	\section{Simulation experiments}\label{sec:sim}
	\subsection{Setup}
	We conduct simulation experiments to assess the performance of the proposed online causal inference framework applied to the G-computation (G-comp), IPTW and AIPTW estimators. We compare our method with the corresponding oracle estimators that operate on the entire pooled dataset.
	The evaluation criteria concern both statistical property and computational efficiency in estimating the average treatment effect $\Delta$: (a) empirical bias (Bias), (b) relative bias, defined as bias divided by the true parameter value (R.bias), (c) average of estimated standard error (ASE), (d) empirical standard error (ESE), (e) coverage probability (CP), 
	(f) the total amount of time required by data loading and algorithm execution (Tol.time) and (g) algorithm execution time only (Run.time). 
	
	Simulation results are summarized over 500 replications. In each replication we generate a total of $b$ batches of data 
	with a total of $N_b$ observations $(X,A,Y)$ under the following data generating mechanism. The measured covariates $X~{\sim}~ \text{MVN}({0},V)$ where $V$ is a compound symmetry covariance matrix with correlation $\rho=0.5$. 
	Conditional on $X$, the treatment indicator $A$ follows a Bernoulli distribution with $\mathbb{P}(A=1\mid X)=[1+\exp\{(1,X\transpose)\alpha\}]\inv$. Conditional on $X$ and $A$, we generate a continuous outcome $Y$ following a Gaussian distribution with mean $\mathbb{E}[Y\mid X, A] =(1,A,X\transpose,AX\transpose)\beta$ and variance one. 
	The true ATE is then derived based on Eq.~\eqref{eq:ate}. 
	All parameters are randomly generated with $\theta=(\alpha\transpose,\beta\transpose,\Delta)\transpose\sim \text{MVN}({0},I_{d})$, where $d$ is the total number of parameters. The simulation code is available at \url{https://github.com/luolsph/OnlineCausal}.

	\subsection{Evaluation of parameter estimation}\label{subsec:sim_est}
	We first consider a typical streaming data setting with fixed batch size $n_j,j=1,\dots,b$ and an increasing number of batches (and hence total sample size $N_b$).  We then study impact of batch size by fixing the total sample size and increasing the batch size.
	
\paragraph{Scenario 1 (streaming data): fixed $n_j$ and increasing  $b$.}
	We consider a streaming data setting where the batch size is fixed at $n_j\equiv 100$ and $b$ increases from $10$ to $10^3$. Correspondingly, the total sample size $N_b$ increases from $10^3$ to $10^5$. 
	Simulation results are summarized in Table~\ref{tab:continuous_2}. 
	As number of data batches $b$ increased, we observed a \textcolor{black}{decreasing trend} in estimation bias in all three online estimators, which confirmed the large sample properties in Theorems~\ref{thm:consist} and~\ref{thm:normal}. 
	Our proposed online methods clearly demonstrated their computation efficiency relative to its offline counterpart as $b$ increases: when processing a large sample with a total of more than $N_b=10^4$ observations, the online methods are roughly two times faster than the offline methods with little loss of estimation precision and inferential power.
	This is expected because the offline methods spent longer time to load in data and to analyze a large polled dataset. 
	
	\begin{table}[h]
		\centering
		\caption{\label{tab:continuous_2}
		Simulation results under scenario 1 (Streaming data setting): number of batches $b$ increases from $10$ and $10^3$, while data batch size is fixed at $n_j\equiv 100$.}
		\begin{tabular}{l rr rr rr}
			\hline
			&\multicolumn{2}{c}{G-comp}  &\multicolumn{2}{c}{IPTW}  &\multicolumn{2}{c}{AIPTW} \\
			&\multicolumn{1}{c}{offline} &\multicolumn{1}{c}{online} &\multicolumn{1}{c}{offline} &\multicolumn{1}{c}{online} 
			&\multicolumn{1}{c}{offline} &\multicolumn{1}{c}{online} \\
			\hline
			&\multicolumn{6}{c}{$b=10$, $n_j=100$, $N_b=1000$} \\
			%	$b$ &1 &10 &50 &1 &10 &50 &1 &10 &50 \\
			\hline
			Bias$\times10^{-3}$ &-4.835 &-4.835 &-7.156 &-7.127 &-7.865 &-6.845 \\
			R.bias$\times10^{-3}$ 
			&-26.96 &-26.96 &-39.90 &-39.74 &-43.86 &-38.17 \\
			ASE$\times10^{-2}$  &7.585 &7.542 &8.212 &8.229 &7.879 &7.860\\
			ESE$\times10^{-2}$  &7.769 &7.769 &8.585 &8.417 &8.077 &8.217\\
			CP   &0.955 &0.955 &0.935 &0.945 &0.950 &0.940\\
			Tol.time (s)  &0.013 &0.021 &0.015 &0.017 &0.054 &0.046\\
			Run.time (s)  &0.012 &0.015 &0.013 &0.011 &0.053 &0.039\\
			\hline
			&\multicolumn{6}{c}{$b=100$, $n_j=100$,  $N_b=10^4$} \\
			\hline
			Bias$\times10^{-3}$ 
			&-0.683 &-0.683 &-1.041 &0.371 &-0.609 &-0.652 \\
			R.bias$\times10^{-3}$ 
			&-3.799 &-3.799 &-5.789 &-2.065 &-3.387 &-3.629 \\
			ASE$\times10^{-2}$  &2.333 &2.331 &2.475 &2.478 &2.384 &2.383 \\
			ESE$\times10^{-2}$  &2.385 &2.385 &2.571 &2.567 &2.459 &2.454 \\
			CP   &0.950 &0.950 &0.938 &0.940 &0.940 &0.944 \\
			Tol.time (s)  &0.108 &0.183 &0.144 &0.128 &0.489 &0.280 \\
			Run.time (s)  &0.097 &0.127 &0.130 &0.075 &0.474 &0.227 \\
			\hline
			&\multicolumn{6}{c}{$b=1000$, $n_j=100$,  $N_b=10^5$} \\
			\hline
			Bias$\times10^{-3}$ 
			&0.056 &0.056 &-0.028 &0.188 &-0.361 &-0.025 \\
			R.bias$\times10^{-3}$ 
			&0.313 &0.313 &-0.153 &1.042 &-0.200 &-0.141 \\
			ASE$\times10^{-2}$  &0.738 &0.738 &0.782 &0.782 &0.754 &0.754\\
			ESE$\times10^{-2}$  &0.682 &0.682 &0.732 &0.733 &0.701 &0.702\\
			CP    &0.968 &0.968 &0.966 &0.968 &0.964 &0.964\\
			Tol.time (s)  &5.439 &1.172 &1.079 &0.817 &4.533 &1.895\\
			Run.time (s)  &5.355 &0.804 &0.999 &0.466 &4.453 &1.508\\
			\hline
		\end{tabular}
	\end{table}

\paragraph{Scenario 2: fixed $N_b$ and varying $n_j$.}
	To evaluate the effect of batch size $n_j$, we fix the total sample size at $N_b=10^4$ and consider two different batch sizes $n_j\equiv 1000$ and $500$. 
	Simulation results are summarized in Table~\ref{tab:sim1_interaction}. 
	Similar to Scenario 1, the online and offline estimators had similar bias and coverage probability. The statistical efficiency was not affected by batch size and depended only on the total sample size $N_b$. The computation efficiency was again confirmed by the relatively shorter computation time of the online methods.  
	
	We further conducted simulation studies when the outcome is binary in Section~A.6 %\ref{appendix:simu} 
	of the Appendix. Similar results were observed except that the IPTW estimator may suffer from a slight instability with binary outcomes.

	\subsection{Evaluation of sequential testing}
	We evaluate the Wald test for $H_0:\Delta=0$ vs. $H_1:\Delta\neq 0$ in terms of  type I error and power with a pre-specified significance level of $\alpha=0.05$. We set a sequence of values for $\Delta$ under $H_1$ from $0.02$ to $0.12$ with an increment of $0.02$. 
We simulate data streams from Gaussian linear model with total sample size $N_b=10^4$ and  batch size $n_j\equiv 1000$. We then compute the empirical type I error and power of the online G-comp, IPTW, and AIPTW estimators from 500 replications.
	
	{As presented in Table~\ref{tab:sim3_seq_test}, the type I error of the Wald test based on the online estimators were very close to the nominal level of 0.05.} The power of G-comp and AIPTW based tests were around the same level at different alternative values of $\Delta$, while the power of tests based on IPTW were steadily lower than the other two. This is expected because when all models are correctly specified, G-comp is essentially a maximum likelihood estimator and AIPTW achieves the semiparametric efficiency bound in the union model where either the PS or outcome model is correctly specified, whereas IPTW is inefficient.

	\begin{table}[h]
		\centering
		\caption{\label{tab:sim1_interaction}
		Simulation results under scenario 2: total sample size is fixed to be $N_b=10^4$, while number of batches $b=10$ or $b=50$.}
	\begin{tabular}{l ccc ccc ccc}
		\hline
		&\multicolumn{3}{c}{G-comp}  &\multicolumn{3}{c}{IPTW}  &\multicolumn{3}{c}{AIPTW} \\
		&\multicolumn{1}{c}{offline} &\multicolumn{2}{c}{online} &\multicolumn{1}{c}{offline} &\multicolumn{2}{c}{online} 
		&\multicolumn{1}{c}{offline} &\multicolumn{2}{c}{online} \\
%		\cline{3-4} \cline{6-7} \cline{9-10}
		$n_j$ & &1000 &200 & &1000 &200 & &1000 &200 \\
		\hline
		Bias$\times10^{-3}$ 
		&-0.683 &-0.683 &-0.683 &-1.041 &-0.321 &0.240 &-0.609 &-0.619 &-0.645\\
		R.bias$\times10^{-3}$ 
		&-3.799 &-3.799 &-3.799 &-5.789 &-1.785 &1.337 &-3.387 &-3.443 &-3.586\\
		ASE$\times 10^{-2}$ &2.333 &2.332 &2.331 &2.475 &2.478 &2.478 &2.384 &2.384 &2.384 \\
		ESE$\times 10^{-2}$ &2.385 &2.385 &2.385 &2.571 &2.564 &2.562 &2.459 &2.460 &2.456 \\
		CP      &0.950 &0.950 &0.950 &0.938 &0.936 &0.940 &0.940 &0.942 &0.944 \\
		Tol.time (s) &0.061 &0.078 &0.118 &0.113 &0.055 &0.087 &0.330 &0.129 &0.207\\
		Run.time (s) &0.054 &0.069 &0.092 &0.099 &0.047 &0.061 &0.315 &0.123 &0.182\\
		\hline
	\end{tabular}
\end{table}

	\begin{table}[h]
		\centering
		\caption{\label{tab:sim3_seq_test}Simulation results for sequential testing with total sample size $N_b=10^4$, number of batches $b=10$, 
		and total  number  of  tests $T=10$. Type I error and power are estimated empirically by simulating data with different values of $\Delta$.}
		\begin{tabular}{l c |rrr rrr}
		\hline
			&{Type I error}  &\multicolumn{6}{c}{Power} \\
			$\Delta-\Delta_0$ &0 &0.02 &0.04 &0.06 &0.08 &0.10 &0.12  \\
			\hline
			G-comp   &0.070 &0.188 &0.514 &0.824 &0.976 &0.996 &1.000  \\
			IPTW     &0.060 &0.178 &0.492 &0.792 &0.958 &0.994 &0.998  \\
			AIPTW    &0.062 &0.174 &0.498 &0.796 &0.970 &0.994 &1.000  \\
		\hline
		\end{tabular}
	\end{table}
	
	\section{COVID-19 vaccine safety monitoring}\label{sec:application}
	We illustrate our proposed methods via a proof-of-concept study of near real-time COVID-19 vaccine safety monitoring using data from the Vaccine Adverse Event Reporting System (VAERS). The VAERS is a national early warning system maintained by the Centers for Disease Control and Prevention (CDC) and the Food and Drug Administration (FDA) to detect possible safety problems in U.S.-licensed vaccines. It is a passive surveillance system: health departments, vaccine manufacturers, and (parents or family members of) vaccine recipients are encouraged to submit a VAERS report when an adverse event occurs after the administration of any vaccine licensed in the United States. As a result, VAERS is subject to reporting bias, tends to underreport numbers of adverse events, lacks denominator data and unbiased comparison group, 
	and is not designed to determine if a vaccine caused a health problem. Nevertheless, VAERS provides CDC and FDA with valuable information on unusual or unexpected patterns of adverse events that necessitate further evaluation.  
	
	COVID-19 vaccination is the largest vaccination campaign in history. From December 14, 2020 to October 5, 2021, more than 400 million doses of COVID-19 vaccines were administered in the United States. Healthcare providers are required to report to VAERS any COVID-19 vaccine administration errors and serious adverse events regardless of whether the vaccine was the cause. De-identified VAERS data are available to the public 4-6 weeks after the report is received and are updated periodically as new reports are received. Therefore, VARES data allow for near real-time detection of potential safety problems post administration of COVID-19 vaccine. 
	
	Our analysis focused on the two most common post-vaccination symptoms reported, which are headache and fatigue, among adult recipients of COVID-19 vaccines between January 1 to September 30, 2021 who reported an adverse symptom to VAERS. We estimated age effect defined as the risk difference comparing the potential outcome had all vaccine recipients been older than 50 to the potential outcome had all vaccine recipients been younger than 50. We applied our proposed online G-computation, IPTW, and AIPTW methods to update the age effect estimates monthly for a total of nine months, adjusting for gender and whether the individual has a life-threatening illness. 
	To validate our results, 
	we also computed the oracle estimates monthly which utilize the largest possible dataset at each month, which pools current month’s data with all previous months’ data. 
	
\begin{figure}[h]
		\centering
		\includegraphics[width=\textwidth]{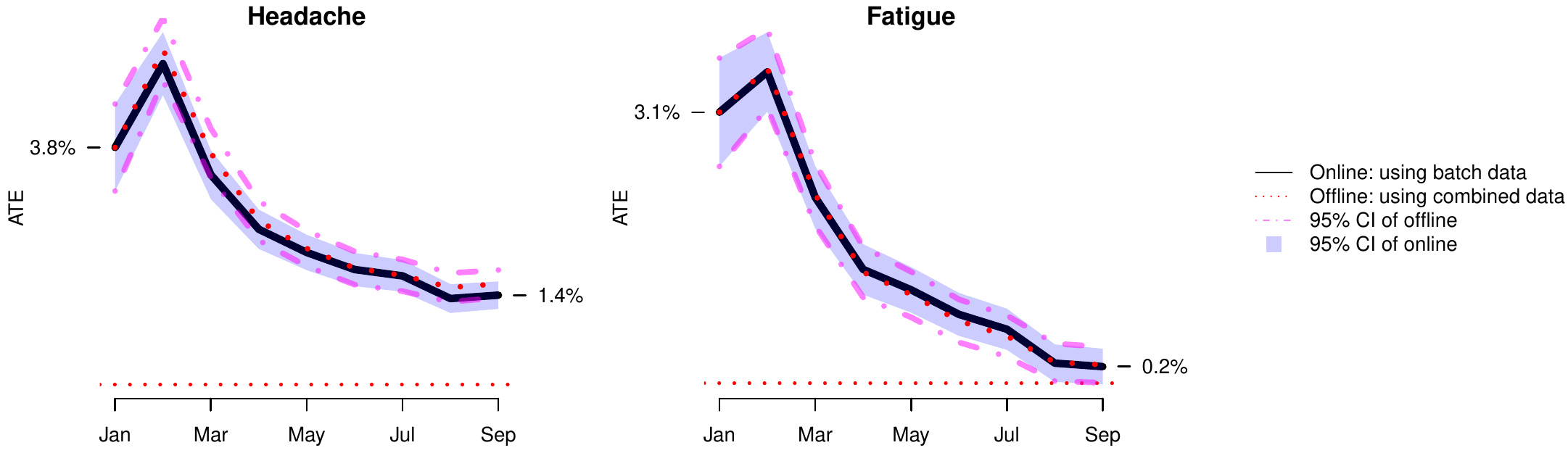}
		\caption{Monthly updates of age effect estimates on headache and fatigue, based on online and oracle offline AIPTW estimators.}
		\label{fig:twooutcomes}
	\end{figure}

	As of September 30, 2021, VAERS received 502,128 reports of COVID-19–associated adverse events among adults, among which 97,212 (19\%) reported headache and 81,125 (16\%) reported fatigue post-vaccination. 
	Figure~\ref{fig:twooutcomes} presents the monthly point estimates and 95\% confidence bands of the AIPTW method based on online causal inference.  
	Among adult recipients of COVID-19 vaccines, we estimated an age effect of 1.43\% (95\% CI: 1.21\%, 1.65\%) in risk of headache and 0.19\% (95\% CI: -0.02\%, 0.39\%) in risk of fatigue, both higher had a vaccine recipient been older than 50 compared to younger. 
Interestingly, for both adverse outcomes we observed a peak in around February. This is expected because at that time COVID-19 vaccine has been prioritized for seniors, individuals with heath conditions, and health care workers. As a result, in the accumulative sample of January and February, older adults tend to be less healthy than younger adults, and the estimated higher risk of adverse events reflected the composition of such a biased sample. Nevertheless, as the study sample accumulated till September, we observed stabilized causal effect estimates from both offline and online methods, a result of Theorem~\ref{thm:diff_EE}. 
	We observed similar results from G-comp and IPTW methods which are presented in Appendix~A.7. %\ref{appendix:fig}.
	
	Our analysis is limited by the quality of VAERS data. First, VAERS data is  subject to reporting bias, hence 
	careful interpretation of the estimated risk differences is required. Electronic health records data from an active safety surveillance system such as the Vaccine Safety Datalink may be a better data source for near real-time monitoring via online causal inference \citep{yih2011active}. 
	Second, VAERS data lacks detailed demographic information, and in our analysis, control of confounding is limited to adjustment for gender and life-threatening illness. Therefore, our analysis is likely subject to unmeasured confounding bias. Despite the limitations, the similar results of our proposed method and the traditional method indicate the promising role of online causal inference in the setting of streaming data.
	
	\section{Discussion}\label{sec:discuss}
	As streaming data becomes pervasive, there is a surge of need for an online causal inference framework that accommodates real-time processing of massive data with high velocity. 
	Conventional offline methods that require full access to the combined raw dataset are time consuming, resource intensive, and may suffer from data storage and sharing barriers. In contrast, our proposed online causal inference framework synchronizes evidence as new data become available, without storing or revisiting prior observations but with the same asymptotic properties as offline methods. Our strategy applies to a wide range of causal inference problems, allows each data batch to be a biased sample, enjoys computation and storage efficiency, and overcomes data sharing barrier when necessary. 
	Our approach has promising application in modern data science and can be extended to accommodate heterogeneous treatment effect, time-varying treatment effect, and multi-center studies, which will be explored in future research.

	\clearpage
	\bibliographystyle{apalike}
	\bibliography{paper-ref}

\begin{thebibliography}{}

\bibitem[Angrist et~al., 1996]{angrist1996identification}
Angrist, J.~D., Imbens, G.~W., and Rubin, D.~B. (1996).
\newblock Identification of causal effects using instrumental variables.
\newblock {\em {Journal of the American Statistical Association}},
  91(434):444--455.

\bibitem[Angrist and Keueger, 1991]{angrist1991does}
Angrist, J.~D. and Keueger, A.~B. (1991).
\newblock Does compulsory school attendance affect schooling and earnings?
\newblock {\em {The Quarterly Journal of Economics}}, 106(4):979--1014.

\bibitem[Bate et~al., 1998]{bate1998bayesian}
Bate, A., Lindquist, M., Edwards, I.~R., Olsson, S., Orre, R., Lansner, A., and
  De~Freitas, R.~M. (1998).
\newblock A bayesian neural network method for adverse drug reaction signal
  generation.
\newblock {\em European journal of clinical pharmacology}, 54(4):315--321.

\bibitem[Bifet et~al., 2015]{Spark2015}
Bifet, A., Maniu, S., Qian, J., Tian, G., He, C., and Fan, W. (2015).
\newblock Streamdm: Advanced data mining in spark streaming.
\newblock In {\em {IEEE} International Conference on Data Mining Workshop,
  {ICDMW} 2015, Atlantic City, NJ, USA, November 14-17, 2015}, pages
  1608--1611.

\bibitem[Brown et~al., 2007]{brown2007early}
Brown, J.~S., Kulldorff, M., Chan, K.~A., Davis, R.~L., Graham, D., Pettus,
  P.~T., Andrade, S.~E., Raebel, M.~A., Herrinton, L., Roblin, D., et~al.
  (2007).
\newblock Early detection of adverse drug events within population-based health
  networks: application of sequential testing methods.
\newblock {\em Pharmacoepidemiology and drug safety}, 16(12):1275--1284.

\bibitem[Brown et~al., 2009]{brown2009early}
Brown, J.~S., Kulldorff, M., Petronis, K.~R., Reynolds, R., Chan, K.~A., Davis,
  R.~L., Graham, D., Andrade, S.~E., Raebel, M.~A., Herrinton, L., et~al.
  (2009).
\newblock Early adverse drug event signal detection within population-based
  health networks using sequential methods: key methodologic considerations.
\newblock {\em Pharmacoepidemiology and drug safety}, 18(3):226--234.

\bibitem[{Chintapalli} et~al., 2016]{Storm_Spark2016}
{Chintapalli}, S., {Dagit}, D., {Evans}, B., {Farivar}, R., {Graves}, T.,
  {Holderbaugh}, M., {Liu}, Z., {Nusbaum}, K., {Patil}, K., {Peng}, B.~J., and
  {Poulosky}, P. (2016).
\newblock Benchmarking streaming computation engines: Storm, flink and spark
  streaming.
\newblock In {\em 2016 IEEE International Parallel and Distributed Processing
  Symposium Workshops (IPDPSW)}, pages 1789--1792.

\bibitem[Cook et~al., 2012]{cook2012statistical}
Cook, A.~J., Tiwari, R.~C., Wellman, R.~D., Heckbert, S.~R., Li, L., Heagerty,
  P., Marsh, T., and Nelson, J.~C. (2012).
\newblock Statistical approaches to group sequential monitoring of postmarket
  safety surveillance data: current state of the art for use in the
  mini-sentinel pilot.
\newblock {\em Pharmacoepidemiology and drug safety}, 21:72--81.

\bibitem[Cui et~al., 2020]{cui2020semiparametric}
Cui, Y., Pu, H., Shi, X., Miao, W., and Tchetgen, E.~T. (2020).
\newblock Semiparametric proximal causal inference.
\newblock {\em arXiv preprint arXiv:2011.08411}.

\bibitem[Dean and Ghemawat, 2004]{MapReduce2004}
Dean, J. and Ghemawat, S. (2004).
\newblock Mapreduce: Simplified data processing on large clusters.
\newblock In {\em OSDI'04: Sixth Symposium on Operating System Design and
  Implementation}, pages 137--150, San Francisco, CA.

\bibitem[Efron, 1993]{efron1993bayes}
Efron, B. (1993).
\newblock Bayes and likelihood calculations from confidence intervals.
\newblock {\em Biometrika}, 80(1):3--26.

\bibitem[Evans et~al., 2001]{evans2001use}
Evans, S.~J., Waller, P.~C., and Davis, S. (2001).
\newblock Use of proportional reporting ratios (prrs) for signal generation
  from spontaneous adverse drug reaction reports.
\newblock {\em Pharmacoepidemiology and drug safety}, 10(6):483--486.

\bibitem[Ghemawat et~al., 2003]{Hadoop2003}
Ghemawat, S., Gobioff, H., and Leung, S.-T. (2003).
\newblock The google file system.
\newblock In {\em Proceedings of the 19th ACM Symposium on Operating Systems
  Principles}, pages 20--43, Bolton Landing, NY.

\bibitem[Godambe, 1960]{godambe1960optimum}
Godambe, V.~P. (1960).
\newblock An optimum property of regular maximum likelihood estimation.
\newblock {\em {Annals of Mathematical Statistics}}, 31(4):1208--1211.

\bibitem[Grigg et~al., 2003]{grigg2003use}
Grigg, O.~A., Farewell, V., and Spiegelhalter, D. (2003).
\newblock Use of risk-adjusted cusum and rsprtcharts for monitoring in medical
  contexts.
\newblock {\em Statistical methods in medical research}, 12(2):147--170.

\bibitem[Hern{\'a}n and Robins, 2006]{hernan2006instruments}
Hern{\'a}n, M.~A. and Robins, J.~M. (2006).
\newblock Instruments for causal inference: an epidemiologist's dream?
\newblock {\em Epidemiology}, pages 360--372.

\bibitem[Huber, 1964]{huber1964robust}
Huber, P.~J. (1964).
\newblock Robust estimation of a location parameter.
\newblock {\em {Annals of Mathematical Statistics}}, 35:73--101.

\bibitem[Huber, 1967]{huber1967under}
Huber, P.~J. (1967).
\newblock The behavior of maximum likelihood estimates under nonstandard
  conditions.
\newblock In {\em Proceedings of the Fifth Berkeley Symposium on Mathematical
  Statistics and Probability: Weather modification}, volume~5, pages 221--233.
  Univ of California Press.

\bibitem[Jennison and Turnbull, 1999]{jennison1999group}
Jennison, C. and Turnbull, B.~W. (1999).
\newblock {\em Group sequential methods with applications to clinical trials}.
\newblock CRC Press.

\bibitem[Lan and Demets, 1983]{Lan1983}
Lan, K. and Demets, D. (1983).
\newblock Discrete sequential boundaries for clinical-trials.
\newblock {\em Biometrika}, 70(3):659 -- 663.

\bibitem[Li et~al., 2011]{li2011propensity}
Li, L., Kulldorff, M., Nelson, J.~C., and Cook, A.~J. (2011).
\newblock A propensity score-enhanced sequential analytic method for
  comparative drug safety surveillance.
\newblock {\em Statistics in Biosciences}, 3(1):45.

\bibitem[Liang and Zeger, 1986]{liang1986longitudinal}
Liang, K.-Y. and Zeger, S.~L. (1986).
\newblock Longitudinal data analysis using generalized linear models.
\newblock {\em Biometrika}, 73(1):13--22.

\bibitem[Lin and Xi, 2011]{lin2011aggregated}
Lin, N. and Xi, R. (2011).
\newblock Aggregated estimating equation estimation.
\newblock {\em Statistics and Its Interface}, 4(1):73--83.

\bibitem[Luo and Song, 2020]{Luo2020}
Luo, L. and Song, P. X.-K. (2020).
\newblock Renewable estimation and incremental inference in generalized linear
  models with streaming datasets.
\newblock {\em Journal of the Royal Statistical Society: Series B}, 82:69--97.

\bibitem[Miao et~al., 2018]{miao2018identifying}
Miao, W., Geng, Z., and Tchetgen~Tchetgen, E.~J. (2018).
\newblock Identifying causal effects with proxy variables of an unmeasured
  confounder.
\newblock {\em Biometrika}, 105(4):987--993.

\bibitem[Moore et~al., 1997]{moore1997reports}
Moore, N., Kreft-Jais, C., Haramburu, F., Noblet, C., Andrejak, M., Ollagnier,
  M., and B{\'e}gaud, B. (1997).
\newblock Reports of hypoglycaemia associated with the use of ace inhibitors
  and other drugs: a case/non-case study in the french pharmacovigilance system
  database.
\newblock {\em British journal of clinical pharmacology}, 44(5):513--518.

\bibitem[Nelson et~al., 2015]{nelson2015methods}
Nelson, J.~C., Cook, A.~J., Yu, O., Zhao, S., Jackson, L.~A., and Psaty, B.~M.
  (2015).
\newblock Methods for observational post-licensure medical product safety
  surveillance.
\newblock {\em Statistical methods in medical research}, 24(2):177--193.

\bibitem[Neyman, 1923]{neyman1923applications}
Neyman, J. (1923).
\newblock Sur les applications de la thar des probabilities aux experiences
  agaricales: Essay des principle. excerpts reprinted (1990) in english.
\newblock {\em Statistical Science}, 5(463-472):4.

\bibitem[O'Brien and Fleming, 1979]{OBrien1979}
O'Brien, P. and Fleming, T. (1979).
\newblock A multiple testing procedure for clinical trials.
\newblock {\em Biometrics}, 35:549 -- 556.

\bibitem[Page, 1954]{page1954continuous}
Page, E.~S. (1954).
\newblock Continuous inspection schemes.
\newblock {\em Biometrika}, 41(1/2):100--115.

\bibitem[Pearl, 2001]{pearl2001proceedings}
Pearl, J. (2001).
\newblock Direct and indirect effects.
\newblock pages 411--420. Morgan Kaufmann, San Francisco (CA).

\bibitem[Pocock, 1982]{Pocock1982}
Pocock, S. (1982).
\newblock Interim analyses for randomized clinical-trials - the group
  sequential approach.
\newblock {\em Biometrics}, 38(1):153 -- 162.

\bibitem[Robbins and Monro, 1951]{Robbins1951}
Robbins, H. and Monro, S. (1951).
\newblock A stochastic approximation method.
\newblock {\em The Annals of Mathematical Statistics}, 22(3):400--407.

\bibitem[Robins and Greenland, 1992]{robins1992identifiability}
Robins, J.~M. and Greenland, S. (1992).
\newblock Identifiability and exchangeability for direct and indirect effects.
\newblock {\em Epidemiology}, pages 143--155.

\bibitem[Rosenbaum and Rubin, 1983]{rosenbaum1983central}
Rosenbaum, P.~R. and Rubin, D.~B. (1983).
\newblock The central role of the propensity score in observational studies for
  causal effects.
\newblock {\em Biometrika}, 70(1):41--55.

\bibitem[Rubin, 1974]{rubin1974estimating}
Rubin, D.~B. (1974).
\newblock Estimating causal effects of treatments in randomized and
  nonrandomized studies.
\newblock {\em Journal of educational Psychology}, 66(5):688.

\bibitem[Sakrison, 1965]{Sakrison1965}
Sakrison, D.~J. (1965).
\newblock Efficient recursive estimation: application to estimating the
  parameter of a covariance function.
\newblock {\em International journal of engineering science}, 3(4):461--483.

\bibitem[Schifano et~al., 2016]{Schifano2016CUEE}
Schifano, E.~D., Wu, J., Wang, C., Yan, J., and Chen, M.-H. (2016).
\newblock Online updating of statistical inference in the big data setting.
\newblock {\em Technometrics}, 58(3):393--403.

\bibitem[Shi et~al., 2020]{shi2020multiply}
Shi, X., Miao, W., Nelson, J.~C., and Tchetgen~Tchetgen, E.~J. (2020).
\newblock Multiply robust causal inference with double-negative control
  adjustment for categorical unmeasured confounding.
\newblock {\em Journal of the Royal Statistical Society: Series B (Statistical
  Methodology)}, 82(2):521--540.

\bibitem[Stefanski and Boos, 2002]{stefanski2002calculus}
Stefanski, L.~A. and Boos, D.~D. (2002).
\newblock The calculus of m-estimation.
\newblock {\em The American Statistician}, 56(1):29--38.

\bibitem[Tchetgen and Shpitser, 2012]{tchetgen2012semiparametric}
Tchetgen, E. J.~T. and Shpitser, I. (2012).
\newblock Semiparametric theory for causal mediation analysis: efficiency
  bounds, multiple robustness, and sensitivity analysis.
\newblock {\em Annals of statistics}, 40(3):1816.

\bibitem[Tchetgen et~al., 2020]{tchetgen2020introduction}
Tchetgen, E. J.~T., Ying, A., Cui, Y., Shi, X., and Miao, W. (2020).
\newblock An introduction to proximal causal learning.
\newblock {\em arXiv preprint arXiv:2009.10982}.

\bibitem[Toulis and Airoldi, 2015]{Toulis2015MLE}
Toulis, P. and Airoldi, E.~M. (2015).
\newblock Scalable estimation strategies based on stochastic approximations:
  classical results and new insights.
\newblock {\em Statistics and computing}, 25(4):781--795.

\bibitem[Wald, 1945]{wald1945sequential}
Wald, A. (1945).
\newblock Sequential tests of statistical hypotheses.
\newblock {\em The annals of mathematical statistics}, 16(2):117--186.

\bibitem[Whitehead, 1997]{whitehead1997design}
Whitehead, J. (1997).
\newblock {\em The design and analysis of sequential clinical trials}.
\newblock John Wiley \& Sons.

\bibitem[Yih et~al., 2011]{yih2011active}
Yih, W.~K., Kulldorff, M., Fireman, B.~H., Shui, I.~M., Lewis, E.~M., Klein,
  N.~P., Baggs, J., Weintraub, E.~S., Belongia, E.~A., Naleway, A., et~al.
  (2011).
\newblock Active surveillance for adverse events: the experience of the vaccine
  safety datalink project.
\newblock {\em Pediatrics}, 127(Supplement 1):S54--S64.

\end{thebibliography}
	
\end{document}